\documentclass[twocolumn]{aastex63}

\newcommand{\Fig}[1]{\hyperref[#1]{Figure~\ref*{#1}}}
\newcommand{\Tab}[1]{\hyperref[#1]{Table~\ref*{#1}}}
\newcommand{\Sec}[1]{\hyperref[#1]{Section~\ref*{#1}}}
\newcommand{\Eq}[1]{\hyperref[#1]{Equation~\ref*{#1}}}

\usepackage{savesym}
\usepackage{amsmath}
\usepackage{hyperref}
\usepackage{booktabs}
\usepackage{comment}
\usepackage{ulem}
\usepackage[version=3]{mhchem}
\usepackage{mathtools}
\usepackage[utf8]{inputenc}
\usepackage{epstopdf}
\usepackage{lineno}

\renewcommand*{\thefootnote}{\fnsymbol{footnote}}

\newcommand{\W}{\color{white}}
\newcommand{\R}{\color{black}}
\def\kms  {km~s$^{-1}$}
\def\masy {mas~yr$^{-1}$~}

\newcommand{\choh}{CH$_3$OH}
\newcommand{\water}{H$_2$O}

\shortauthors{Hyland et al.}
\shorttitle{Perseus-Sagittarius Arm Merge}
\submitjournal{ApJ}

\begin{document}
	\title{An updated model for the Perseus Spiral Arm from Trigonometric Parallax and 3-dimensional kinematic distances of distant massive stars}
    
	\correspondingauthor{Lucas Jordan Hyland}
	\email{Lucas.Hyland@utas.edu.au}
	\author[0000-0002-4783-6679]{L. J. Hyland}
	\affil{School of Natural Sciences, University of Tasmania, Private Bag 37, Hobart, Tasmania 7001, Australia}
	\affil{International Centre for Radio Astronomy Research, The University of Western Australia, 35 Stirling Highway, Crawley WA 6009, Australia}
	\author[0000-0001-7223-754X]{M. J. Reid}
	\affil{Center for Astrophysics $\vert$ Harvard and Smithsonian, 60 Garden Street, Cambridge, MA 02138, USA}
	\author[0000-0002-1363-5457]{S. P. Ellingsen}
	\affil{International Centre for Radio Astronomy Research, The University of Western Australia, 35 Stirling Highway, Crawley WA 6009, Australia}
	\author[0000-0003-4468-761X]{A. Brunthaler}
	\affiliation{Max-Planck-Institut f$\ddot{u}$r Radioastronomie, Auf dem H{\" u}gel 69, 53121 Bonn, Germany}
	\author{X. W. Zheng}
	\affiliation{Nanjing University, Nanjing 20093, China}
    \author[0000-0001-6459-0669]{K. M. Menten$^\ddagger$}\footnote{Sadly, Prof Karl M. Menten passed away before the publication of this manuscript}
	\affiliation{Max-Planck-Institut f$\ddot{u}$r Radioastronomie, Auf dem H{\" u}gel 69, 53121 Bonn, Germany}

\renewcommand*{\thefootnote}{\arabic{footnote}}

\begin{abstract}
    We report trigonometric parallaxes and proper motions for three water masers and one methanol maser obtained with the VLBA as part of the BeSSeL Survey. Incorporating these parallaxes with 3-dimensional kinematic distances, we refine the position and pitch angle of the  Perseus spiral arm in the 1$^\mathrm{st}$ Galactic quadrant. Extrapolating the Perseus and Sagittarius arm locations to beyond the Galactic Center, we find that they intersect at an approximate Galactocentric azimuth of ${200^\circ}$ and radius of $5.6$~kpc.
  \end{abstract}	
	
\section{Introduction}
    We are still refining our understanding of the size and structure of the Milky Way. The Bar and Spiral Structure Legacy (BeSSeL) Survey \citep{Brunthaler2011AN....332..461B, Reid2009ApJ...693..397R, Reid2014ApJ...783..130R, Reid2019ApJ...885..131R}, and the VLBI Exploration of Radio Astrometry \citep[VERA; ][]{Honma2012PASJ...64..136H, Vera2020PASJ...72...50V} project have measured over 250 parallaxes to masers associated with high-mass star-forming regions. These measurements locate spiral arms in the $1^\mathrm{st},2^\mathrm{nd}$ and $3^\mathrm{rd}$ Galactic quadrants out to a distance from the Sun of $\approx10$~kpc. Observations by Southern Hemisphere  Parallax Interferometric Radio Astrometry Legacy Survey \citep[S$\pi$RALS; ][]{Hyland2023ApJ...953...21H} are now providing parallaxes in the $4^\mathrm{th}$ quadrant with distances expected up to 10~kpc \citep{Hyland2025arXiv251016998H}. The final frontier of Galactic VLBI astrometry is to measure accurate distances to sources located on the far side of the Milky Way and through these improve our knowledge of the associated spiral arms and other structures. 
    
    The Perseus spiral arm has long been considered one of the major arms of the Milky Way \citep[e.g., ][]{Georgelin1976A&A....49...57G},
    but its structure in distant regions past the Galactic center remains uncertain.  Recently, \citet{Xu2023ApJ...947...54X} and \citet{Bian2024AJ....167..267B} presented evidence that it merges with the Sagittarius-Carina arm in this distant part of the Milky Way. In this paper, we present Very Long Baseline Array (VLBA\footnote{The VLBA is operated by the National Radio Astronomy Observatory. The National Radio Astronomy Observatory is a facility of the National Science Foundation operated under cooperative agreement by Associated Universities, Inc.}) astrometric measurements for four astrophysical masers in the $1^{\rm st}$ quadrant of the Perseus arm. Using these new measurements and by incorporating 3-dimensional kinematic distances \citep[3DKD; ][]{Reid2022AJ....164..133R} we are able to accurately determine distances to these masers. We model the location and shape of the Perseus spiral arm over nearly its full trace and estimate where it appears to merge with the Sagittarius-Carina arm.

\section{Methods} 
    The absolute distance accuracy for maser VLBI trigonometric parallax measurements has been recorded as good as 1.2~pc \citep[to the Pleiades cluster at 136.2~pc; ][]{Melis2014Sci...345.1029M}, however, due to the inverse relationship between parallax and distance, sources at large distances have a much poorer absolute accuracy (as distance uncertainty is proportional to distance-squared) and are subject to a bias from an asymmetric probability distribution, making the determination of distant portions of the spiral arms difficult.

    
    Standard kinematic distances are subject to significant issues, namely, for Galactic radii ($R$) smaller than the Sun's Galactic radius ($R_0$), there are two distances that have the same line-of-sight velocity relative to the local standard of rest, $V_\mathrm{lsr}$. This can also be considered probabilistically -- the measured velocity (with uncertainty) in the inner Galaxy will lead to two equal peaks in the distance probability distribution function (PDF). Additional measurements can be used to select which peak is more likely, for example, by incorporating proper motions \citep{Sofue2011PASJ...63..813S}. 
    
    Proper motions are the relative angular movement of the target with respect to the Sun, tangential to the line of sight. Scaling proper motions by distance gives velocities, e.g., the proper motion in Galactic longitude $\mu_{l*}$ (with units mas~yr$^{-1}$) is converted to a velocity $v_{l*}$ (in \kms) by multiplying by the distance $d$ (in kpc): $v_{l*}= 4.74~d~\mu_{l*}$. Using a rotation curve to model $\mu_{l*}$ at different sampled distances, one can construct a second PDF describing the likelihood of measuring $\mu_{l*}$ if the target is at a given distance. This method improves upon solely using the $V_\mathrm{lsr}$ kinematic distance and can resolve mbiguities. 
    Given a line of sight $(l,b)$ (where $l$ is the Galactic longitude, $b$ is the Galactic latitude) an adopted Galactic rotation curve, and measurements of the line-of-sight velocity $V_\mathrm{lsr}$ and transverse proper motions $\mu_{l*}$ and $\mu_b$, the 3DKD method uses these kinematic constraints to build a set of likelihoods from which the maximum likelihood distance is inferred \citep{Reid2022AJ....164..133R}.
    
    A key point is that, while the distance uncertainty of trigonometric parallaxes inflates with distance, 3DKD uncertainty is consistent at approximately 1.5~kpc for the current VLBA data \citep[for $d>5$~kpc, see ][]{Reid2022AJ....164..133R}. The primary limitations are how well the Milky Way rotation curve is known (uncertainty $<5$~\kms~ for $R>5$~kpc), and maser source non-circular motions \citep[$\approx7$~\kms~per coordinate; ][]{Reid2019ApJ...885..131R}. In practice, with current VLBA astrometric capabilities, trigonometric parallaxes generally yield more accurate distance estimates for sources $\lesssim8$ kpc from the Sun, while 3DKDs are often more accurate for more distant sources \citep{Yamauchi2016PASJ...68...60Y,Reid2022AJ....164..133R,Bian2024AJ....167..267B}. By incorporating both 3DKD and parallax, we can more accurately determine distances. 
    
\section{Observations and Calibration} \label{sec:obs} 
    \begin{table*}[ht]
		\footnotesize
		\centering
		\caption{Information on observed masers and quasars. Columns: \textbf{(1)} Name in Galactic coordinates for masers and J2000 for quasars, \textbf{(2)} Right Ascension in J2000, \textbf{(3)} Declination in J2000, \textbf{(4)} average flux density: in self-calibrated images, integrated over the emission region for quasars, in spectrum for masers, \textbf{(5)} separation between maser and quasar, \textbf{(6)} velocity relative to the local standard of rest, \textbf{(7)} maser type-- either 22~GHz \water\ or 6.7~GHz class II \choh, \textbf{(8)} reference for maser discovery 1: \citet{GenzelDownes1979A&A....72..234G}, 2: \citet{Codella1995MNRAS.276...57C}, 3: \citet{Szymczak2000A&AS..143..269S}, 4: \citet{GenzelDownes1977A&AS...30..145G}. }
		\label{tab:br210_targets}
		\begin{tabular}{rccccccc}
			\toprule
			\textbf{Name}&\textbf{R.A J2000}&\textbf{DEC J2000}&$\boldsymbol{S}$&$\boldsymbol{\theta_\mathrm{sep}}$&$\boldsymbol{V_\mathrm{lsr}}$&{\bf Type}&{\bf Ref}\\
			&{\bf(h~~m~~s)}&{\bf(d~~m~~s)}&{\bf(Jy)}&{\bf(deg)}&{\bf(\kms)}&&\\
			\midrule
	       G021.87$+$00.01  & 18 31 01.7367 &--09 49 01.116 & $\lesssim1$               & ...   &$+19.6$ & \water& 1\\
			J1825$-$0737  & 18 25 37.6096 &--07 37 30.013 & $0.120$   & 2.566 &  ...   &       \\        
			J1835$-$1115  & 18 35 19.5754 &--11 15 59.326 & $<0.010$             & 1.793 &  ...   &       \\\hline 
           G037.82$+$00.41  & 18 58 53.8794 & +04 32 15.004 & $\approx15$               & ...   &$+19.0$ & \water& 2\\
			J1855$+$0251  & 18 55 35.4364 & +02 51 19.563 & $0.070$  & 1.874 &  ...   &       \\
			J1856$+$0610  & 18 56 31.8388 & +06 10 16.765 & $0.150$ & 1.738 &  ...   &       \\\hline
          G060.57$-$00.18  & 19 45 52.5019 & +24 17 42.749 & $\approx5$               & ...   &$+3.7$  & \choh & 3\\
			J1946$+$2418  & 19 46 19.9607 & +24 18 56.909 & $0.025$    & 0.116 &  ...   &       \\
			J1949$+$2421  & 19 49 33.1420 & +24 21 18.245 & $0.125$  & 0.921 &  ...   &       \\\hline 
	       G070.29$+$01.60  & 20 01 45.3486 & +33 32 45.711 & $\approx1$               & ...   &$-23.3$ & \water& 4\\
			J1957$+$3338  & 19 57 40.5499 & +33 38 27.943 & $0.125$  & 1.024 &  ...   &       \\
			J2001$+$3323  & 20 01 42.2090 & +33 23 44.765 & $0.135$ & 0.151 &  ...   &       \\
			\bottomrule
		\end{tabular}
		\vspace{0.5cm}
	\end{table*}
    We conducted astrometric observations for three 22~GHz water masers, G021.87+00.01 (G021), G037.82+00.41 (G037), and G070.29+01.60 (G070), and one 6.7~GHz methanol maser, G060.57--00.18 (G060), with the VLBA between 2016–2017. The targets and their respective calibrators are listed in Table~\ref{tab:br210_targets}. Each of the selected targets was a candidate for Perseus arm membership based on $(l,b,V_\mathrm{lsr})$. 
       
    The data were collected as part of the BeSSeL Survey BR210 program, and the data are accessible and downloadable from the NRAO archive. This program included many masers thought to be distant and were each observed in three sessions of 4-8-4 epochs respectively, sampled near the Right Ascension peaks of the parallax curve. Each season was separated by 6~months, and epochs within each season were separated by 1-2~week for the middle season, and 3-4 weeks for the other two. Within each epoch, target masers and quasars were observed in a nodding phase referencing mode, with a switching time of 30~seconds. 

    The telescope baseband data were correlated with the DiFX software correlator \citep{Deller2011PASP..123..275D} at Socorro, New Mexico, and reduced in NRAO AIPS \citep{Greisen2003ASSL..285..109G} and ParselTongue \citep{Kettenis2006ASPC..351..497K} using a now-standardized VLBI astrometry process. As this has been discussed in many other BeSSeL publications, we will omit the details but refer the reader to \citet{Reid2009ApJ...693..397R}.
    
    In each epoch, the final products from this reduction are amplitude- and delay-calibrated visibilities, which have been phase referenced from each of the quasars to the maser, or from a single maser channel to both quasars (inverse phase reference). This subtracts the time-variable atmospheric phase along the line of sight \citep{Alef1989ASIC..283..261A, BC1995ASPC...82..327B}, leaving the relative position difference to be measured in the resulting synthesized images. 

    For the 6.7~GHz methanol maser G060, we used inverse phase referencing to image each of the two quasar calibrators in all 16 epochs. For the water masers, G021 could only be referenced against a single quasar (as the other quasar was too weak in every epoch). Additionally, for the quasar that could be used, the maser could only be found in synthesized images in 8 of 16 epochs due to maser variability. G037 was referenced against both quasars in all 16 epochs, and G070 was referenced against both quasars in 16 and 14 epochs, respectively (Figure~\ref{fig:per_parallaxes}). The measured positions at each epoch for each maser are included in the online supplementary material, with columns giving epoch, East-West offset and uncertainty, and North-South offset and uncertainty.
    
\section{Analysis} 
	\subsection{Trigonometric Parallax and Proper Motion}
\begin{figure*}[ht]
    		\centering
            \includegraphics[width=.49\textwidth]{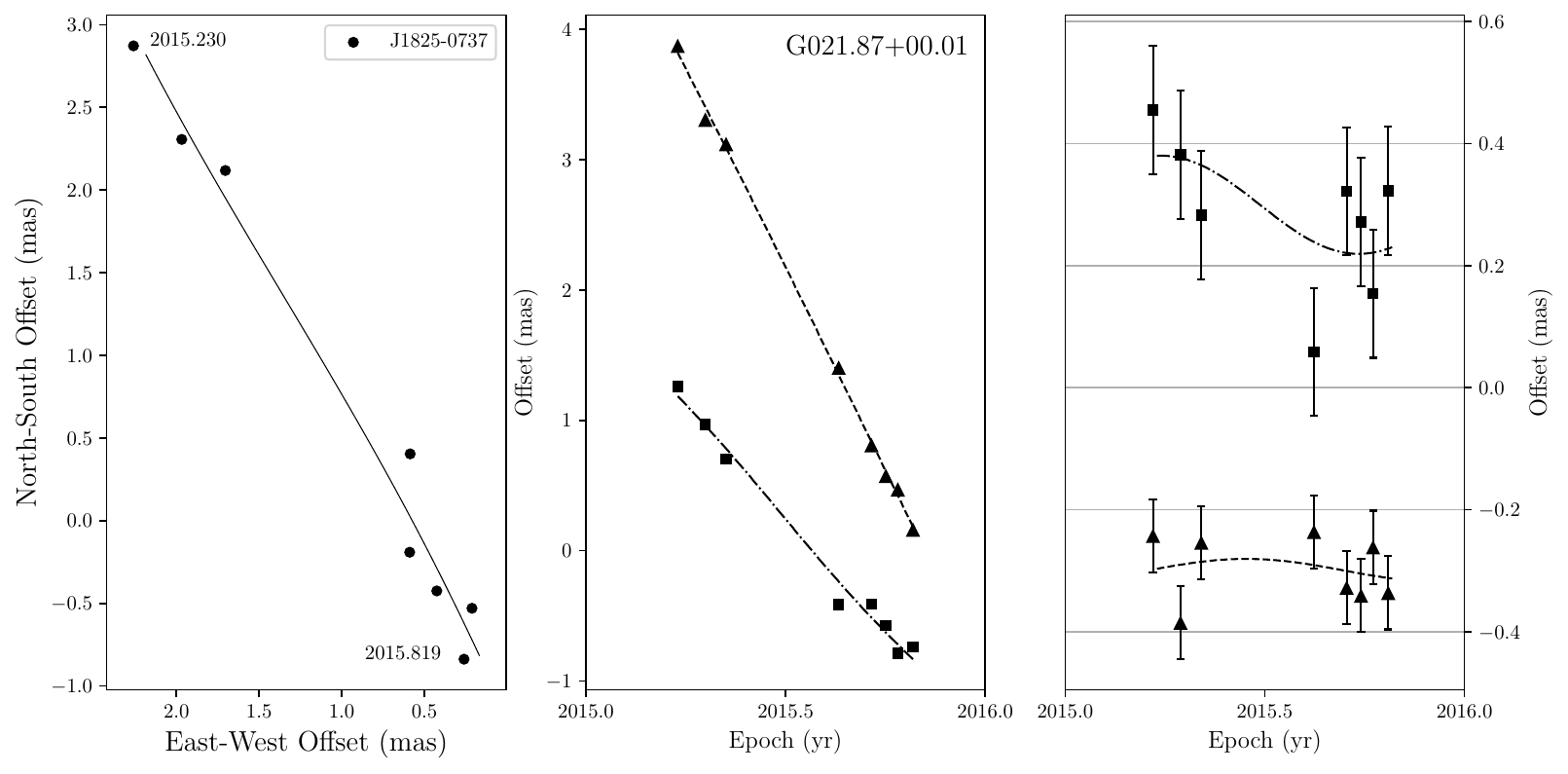}
            \includegraphics[width=.49\textwidth]{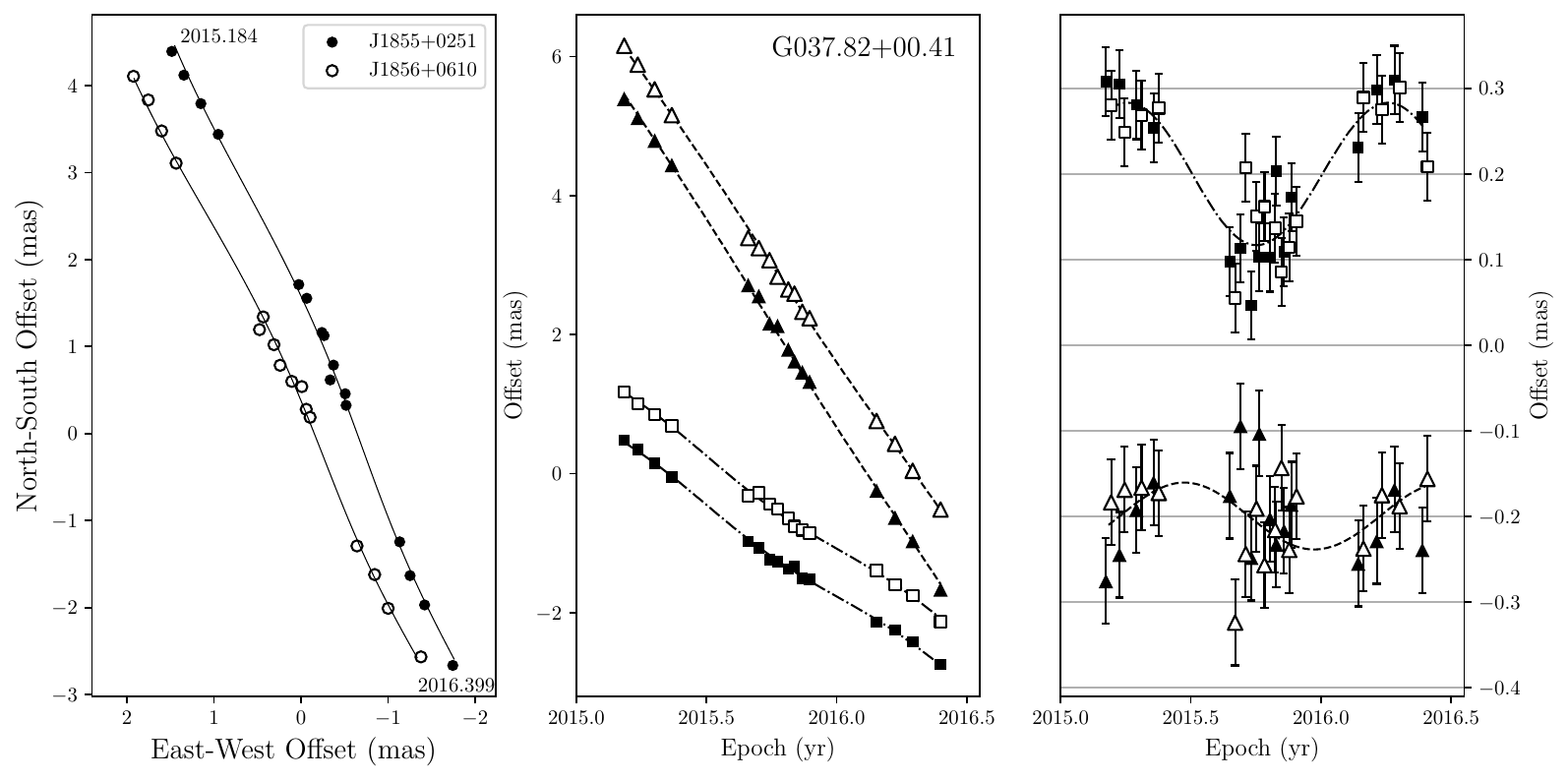}
            ~
            \includegraphics[width=.49\textwidth]{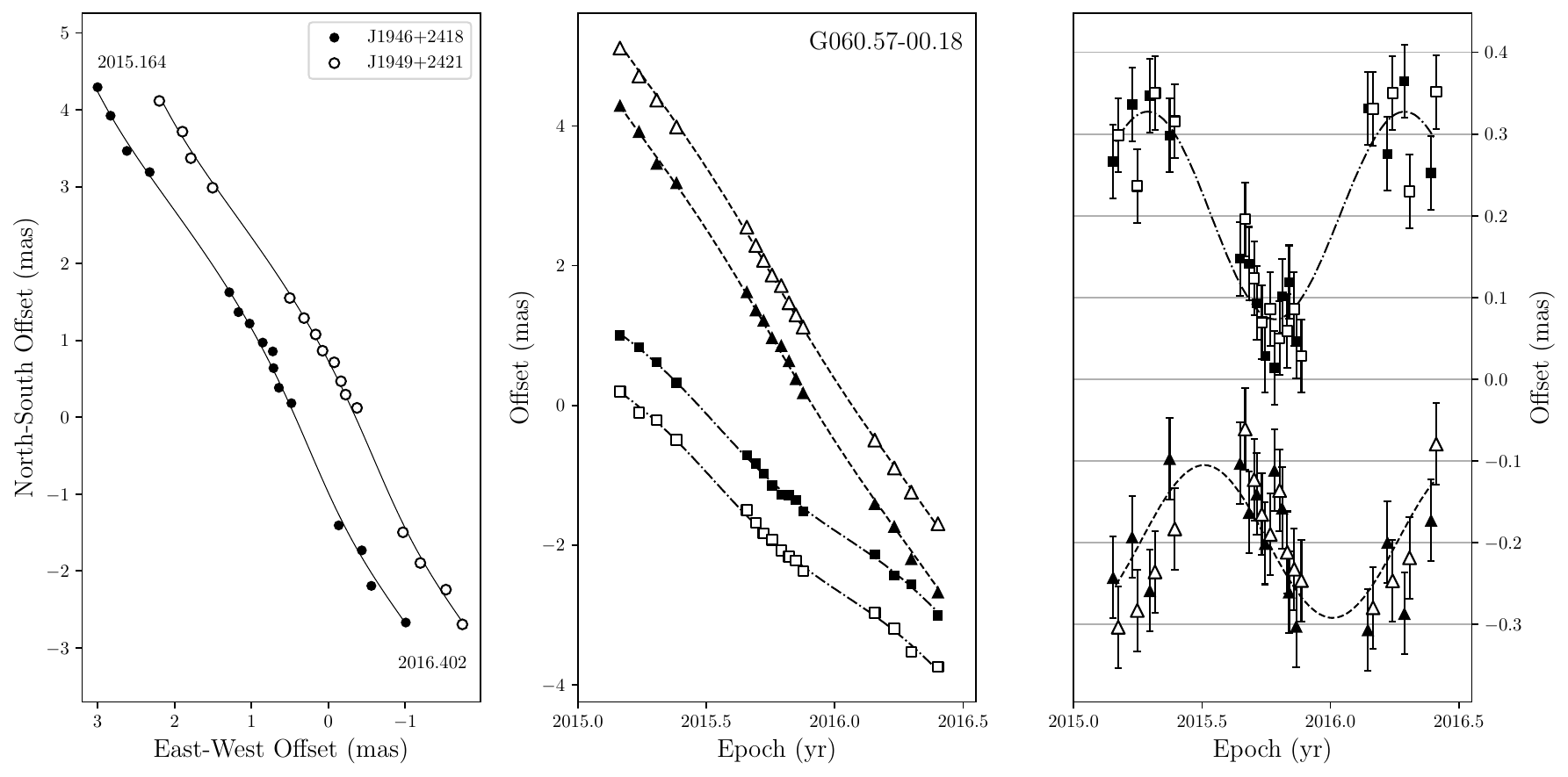}	
    		\includegraphics[width=.49\textwidth]{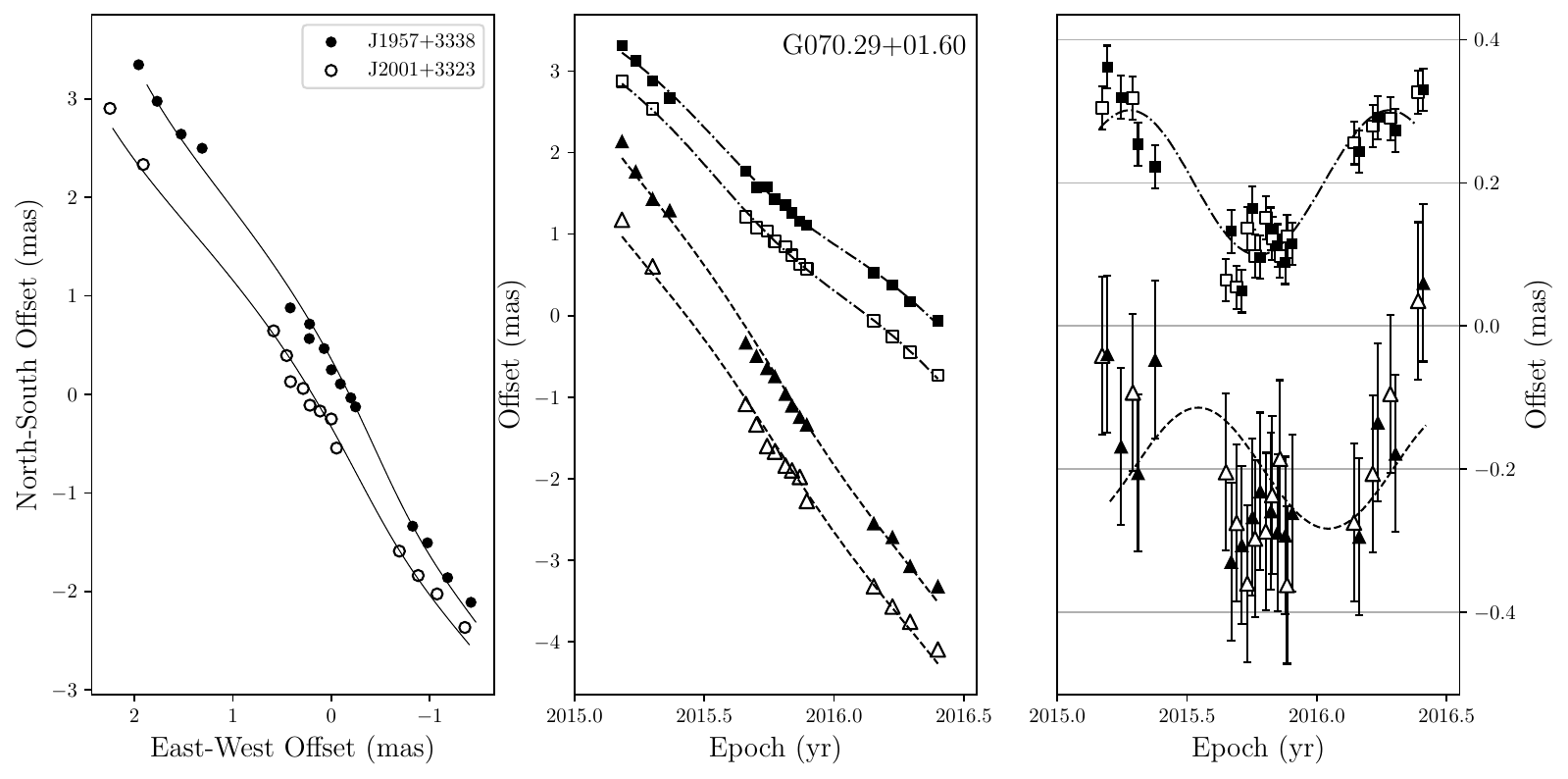}
    		\caption{Parallaxes and proper motion fits for G021.87$+$00.01, G037.82$+$00.41, G060.57--00.18, and G070.29$+$01.60. Filled and open symbols represent motion with respect to different reference quasars in the order they appear in Table 1. {\bf Left panels:} Total sky motion, the y-axis is North-South offsets, and the x-axis is East-West offsets. {\bf Center panels:} Time--varying positions in North--South (triangles) and East--West (squares) directions vs. time. The x-axis is time in years, the y-axis is the offset in the corresponding direction. {\bf Right panels:} Proper motion-subtracted time-varying positions in aforementioned directions. For clarity, RA and declination signatures for each reference quasar have been offset from one another.}
    		\label{fig:per_parallaxes}
    \end{figure*}

    We modeled the measured positions $(x_m, y_m)$\footnote{$x=\Delta\alpha\cos\delta, y=\Delta\delta$} relative to a quasar as a function of time using a least-squares fit that simultaneously solves for the trigonometric parallax ($\pi$), proper motions in the East-West ($\mu_x$) and North-South ($\mu_y$) direction, and position offsets near the center of the data $(x_0, y_0)$ \citep[following ][]{Reid2009ApJ...693..397R}. The parallax signature was computed using Earth–Sun vectors from the NOVAS astrometric routines \citep{Kaplan1989AJ.....97.1197K}. The fits are shown in Figure~\ref{fig:per_parallaxes}, with the results tabulated in Table~\ref{tab:br210_parallax}. 
    
    There were insufficient data to measure an accurate trigonometric parallax for G021, as there were no persistent maser spots to track the motion over the full year, and there were no maser detections in the first two, two out of the middle eight, and in any of the final four epochs. The primary consequence of this is that the trigonometric parallax and $\mu_x$ proper motion parameters become strongly correlated in the model fit, resulting in poorly constrained estimates for both quantities. To break this degeneracy, we first fit the data for ($\pi,\mu_x,\mu_y$), then use the proper motions (and line-of-sight velocity) to calculate a 3D kinematic distance ($d_\mathrm{3Dk}$, as described in Section~\ref{sec:dist}). We then re-fit ($\mu_x,\mu_y$), with $\pi$ fixed at $1/d_\mathrm{3Dk}=0.073$~mas. A subsequent iteration produced no significant change in the fitted parameters, indicating convergence. Both the ``raw'' results and final 3D kinematic-dependent results are presented in Table~\ref{tab:br210_parallax}.

    We present $x,y$ proper motions in Table~\ref{tab:br210_parallax} with their \textit{formal} uncertainties for the maser spots as determined by the fitting process. However, we ultimately want the proper motion and $V_\mathrm{lsr}$ of the central exciting star for Galactic kinematics, which can differ from that measured for a given maser spot owing to internal source motions. First, we look for a peak in the CO \citep[i.e., from ][]{Dame2001ApJ...547..792D} that we can assign as the likely bulk recession velocity of the molecular cloud, and use that as $V_\mathrm{lsr}$ moving forward. 
    
    Methanol masers are known to have a relatively low relative velocity compared to the star \citep{Moscadelli2010ApJ...716.1356M} and generally have a velocity range $<15$~\kms~ \citep{Green2017MNRAS.469.1383G}. G060 has only a single maser spot, so we conservatively add 5~\kms (which, at the measured distance of 7.7~kpc, gives 0.13~\masy) in quadrature to the proper motion and $V_\mathrm{lsr}$ uncertainties for G060. 
    Our water masers G021, G037, and G070 have $V_\mathrm{lsr}$ ranges of 6, 10, and 20~\kms~ respectively and we add 10~\kms~ in quadrature to the proper motion and $V_\mathrm{lsr}$ uncertainties. These values and uncertainties are given in Table~\ref{tab:allmasers}.
    
    Finally, we rotated the proper motions from R.A/DEC into the Galactic reference frame $\mu_x,\mu_y\rightarrow\mu_{l*},\mu_b$ (where $\mu_{l*}=\mu_l\cos b$). 

    \begin{table*}[ht]
		\footnotesize
		\centering
		\caption{Measured parallax and proper motions of masers. \textbf{(1)} Maser name, \textbf{(2)} quasar name, \textbf{(3)} measured parallax, \textbf{(4)} measured proper motion in East--West direction, \textbf{(5)} measured proper motion in North--South direction. Sources used as the reference in each case are indicated with $^*$. Uncertainties are given as $\pm1\sigma$ for parallaxes and proper motions. \textit{Var-wgt avg} is the variance--weighted average of the previous two rows. For G021.87+00.01, the second row gives the proper motions refitted with the parallax fixed at the inverse of the 3DKD (see Section 4.1).}
		\label{tab:br210_parallax}
		\begin{tabular}{rcccc}
			\toprule
			\textbf{Maser}&\textbf{Reference}&$\boldsymbol{\pi}$&$\boldsymbol{\mu_x}$&$\boldsymbol{\mu_y}$\\
			\textbf{Name}&\textbf{Quasar}&{\bf(mas)}&{\bf(mas\,yr$\boldsymbol{^{-1}}$)}&{\bf(mas\,yr$\boldsymbol{^{-1}}$)}\\\midrule
			G021.87$+$00.01 &J$1825-0737^*$      &  $0.177\pm0.167$ & $-2.79\pm0.69$ & $-6.10\pm0.12$ \\
            \multicolumn{2}{l}{Using 3DKD}        &  $0.073\pm0.006$ & $-3.17\pm0.17$ & $-6.14\pm0.09$ \\\hline             
			G037.82$+$00.41 &J$1855+0251^*$  &  $0.093\pm0.010$ & $-2.622\pm0.025$ & $-5.837\pm0.036$   \\
			                &J$1856+0610^*$  &  $0.074\pm0.011$ & $-2.660\pm0.028$ & $-5.514\pm0.037$    \\
			\textit{Var-wgt avg}&            &  $0.084\pm0.007$ & $-2.639\pm0.019$ & $-5.680\pm0.025$  \\\hline 
			G060.57--00.18$^*$  &J$1946+2418$   &  $0.130\pm0.011$ & $-3.237\pm0.028$ & $-5.729\pm0.040$    \\
			                    &J$1949+2421$   &  $0.131\pm0.014$ & $-3.217\pm0.036$ & $-5.638\pm0.033$   \\
			          \textit{Var-wgt avg}&       &  $0.130\pm0.009$ & $-3.229\pm0.022$ & $-5.675\pm0.025$ \\\hline 
            G070.29$+$01.60 &J$1957+3338^*$& $0.099\pm0.013$ & $-2.748\pm0.032$ &$-4.566\pm0.078$  \\
                            &J$2001+3323^*$& $0.109\pm0.011$ & $-2.984\pm0.029$ &$-4.394\pm0.093$  \\
            \textit{Var-wgt avg}&           & $0.104\pm0.008$ & $-2.877\pm0.021$ &$-4.444\pm0.042$ \\\bottomrule
		\end{tabular}
    \end{table*}

    \subsection{Perseus arm candidate masers} 
    Beginning with the list of all known masers with parallaxes and/or proper motions \citep[][and references therein]{Reid2019ApJ...885..131R,Vera2020PASJ...72...50V,Sakai2022PASJ...74..209S}, we searched for masers associated with the Perseus arm. For $l>10^\circ$, we matched maser positions to the expected $l,b,V_\mathrm{lsr}$ trace known from CO measurements \citep{Dame2001ApJ...547..792D, Reid2016ApJ...823...77R}, without using the parallax or proper motions. This gave a list of masers similar to those discussed in (and primarily from) \citet{Choi2014ApJ...790...99C,Reid2019ApJ...885..131R,Sakai2019ApJ...876...30S,Sakai2022PASJ...74..209S,Zhang2013ApJ...775...79Z, Zhang2019AJ....157..200Z}. For masers within $l\leq10^\circ$, where many spiral arm traces coincide in $l-v$ space, we used the parallax and proper motion measurements to calculate maximum likelihood distances (see Section~\ref{sec:dist}), and identified one more Perseus candidate maser G001.00-00.23 \citep{Reid2019ApJ...885..131R}. Two other masers, G229.57+00.15 and G240.31+00.07 \citep{Choi2014ApJ...790...99C}, which were previously considered as part of the arm, are a little more ambiguous. We will include G229.57, as it has a $V_\mathrm{lsr}=53\pm10$~\kms, which is  consistent with the Perseus arm ($V_\mathrm{lsr}=-53$~\kms), and exclude G240.31, as it does not ($V_\mathrm{lsr}=-68\pm10$~\kms). Altogether we assembled 47 candidate masers (see Table~\ref{tab:allmasers}). 
     
    \subsection{Distance determination} \label{sec:dist}
    Using the astrometric ($l,b,\pi$) and kinematic ($V_\mathrm{lsr},\mu_{l*},\mu_b$) parameters, we constructed PDFs for distance: $P_\pi$ from parallax, $P_{V_\mathrm{lsr}}$ from Doppler velocity, $P_{\mu_{l*}}$ from longitude motion, and $P_{\mu_b}$ from latitude motion. We include a brief description of the process in the Appendix. Due to potential non-circular motions and as the non-parallax PDFs rely on models rather than direct measurement alone, to be conservative, we allow for possible systematics by adding a small constant probability density (of 0.0005~kpc$^{-1}$) to each 0.025~kpc bin of those PDFs. 
    
\begin{figure}
        \centering
        \includegraphics[width=0.95\linewidth]{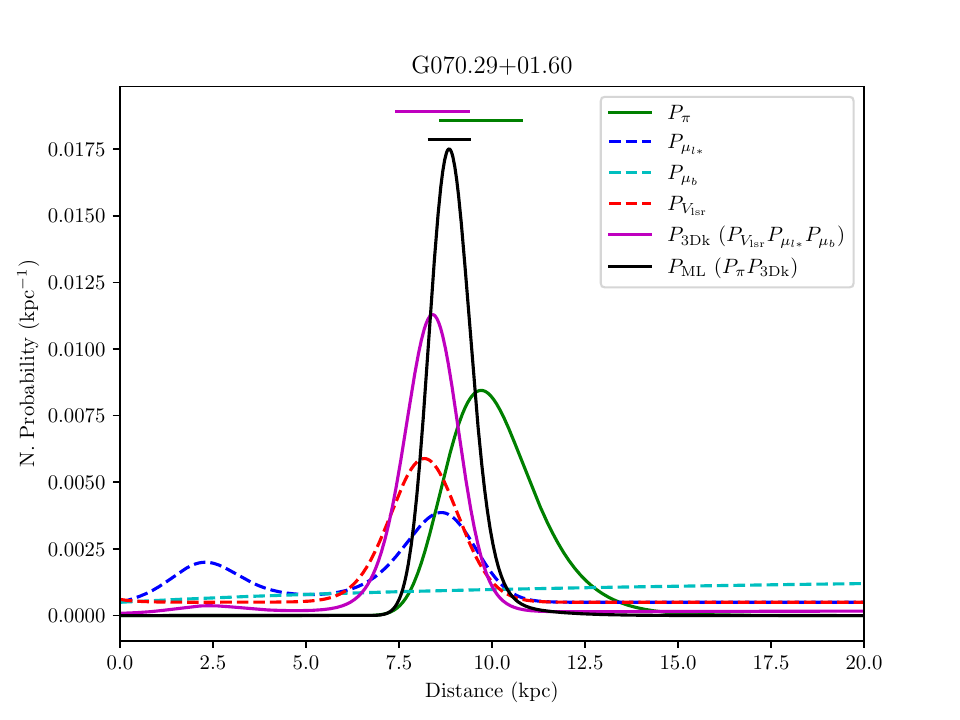}
        \caption{Example of the distance determination process for G070.29+01.60. The $V_\mathrm{lsr}$ PDF (dashed-red) and $\mu_{l*},\mu_b$ (dashed-blue, cyan) proper motion PDFs are multiplied together to give the 3D kinematic distance PDF (magenta). This is then multiplied by the parallax PDF (green) to give the final PDF (black), the peak of which gives the maximum likelihood distance.}
        \label{fig:g070_bayes}
    \end{figure}

    The parallax distance is estimated by finding the median of $P_\pi$, $d=1/\pi_\mathrm{meas}$. The 3D kinematic distance is determined via the method of \citet{Reid2022AJ....164..133R} -- multiplying the $P_{V_\mathrm{lsr}}, P_{\mu_{l*}}, P_{\mu_{b}}$ PDFs and searching for a peak in the new PDF ($P_\mathrm{3Dk}$). Finally, by multiplying $P_\pi$ and $P_\mathrm{3Dk}$, the peak of the resulting PDF ($P_\mathrm{ML}$) gives the maximum likelihood distance ($d_\mathrm{ML}$). Figure~\ref{fig:g070_bayes} gives a graphical example of this process.

    Many of the PDFs we are dealing with are asymmetric, multi-peaked, or otherwise non-Gaussian. We therefore estimate uncertainties by determining the smallest contiguous distance interval of the respective PDF that when integrated, contains 68.3\% of the total probability. The bounds of this interval are quoted as the lower and upper uncertainties on the peak value in Table~\ref{tab:allmasers}.
    
    \citet{Bian2024AJ....167..267B} calculate final distances by taking the variance-weighted average of the parallax distance and 3DKD, where the uncertainty is the inverse square root sum of the weights. Although this does not account for asymmetric PDFs and may underestimate the uncertainties in those cases, their approach mostly gives similar results since the vast majority of parallaxes have less than 15\% uncertainties.

    Some masers in the $2^{\rm nd}$ quadrant of the Perseus arm exhibit non-circular motions exceeding 20~\kms~ \citep{Sakai2019ApJ...876...30S}. This leads to anomalous kinematic distances \citep{Xu2006Sci...311...54X}; consequently, we do not incorporate standard or 3DKD estimates (relying instead on purely the parallaxes) for seven masers\footnote{G108.42+00.89, G108.47-02.81, G108.59+00.49, G111.54+00.77, G122.01-07.08, G133.94+01.06, and G134.62-02.19}, to avoid introducing a known additional systematic error.
   
\begin{figure}
        \centering
        \includegraphics[width=0.99\linewidth]{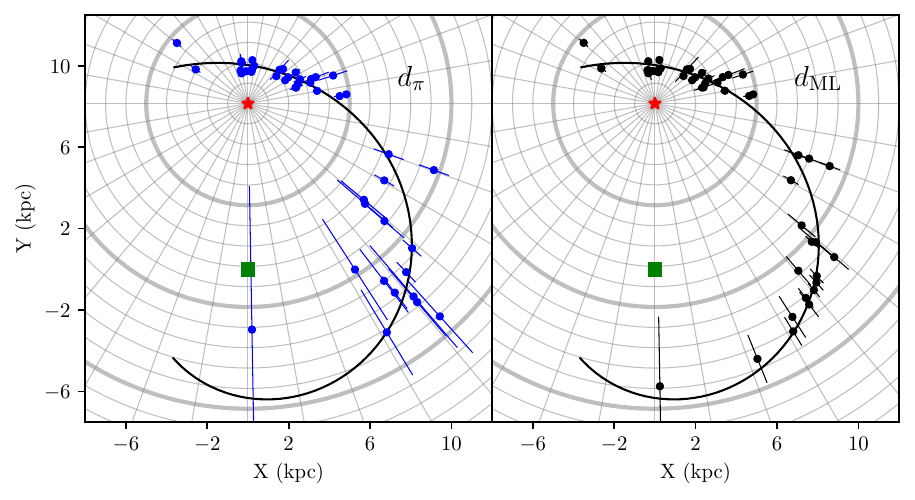}
        \caption{Comparison of parallax distance with maximum likelihood distance in the Galactocentric coordinates frame. \textbf{Left}: Parallax-only distances (blue). \textbf{Right:} Maximum likelihood distances, based on parallax and 3DKD (black). The solid line is the single-segment arm fit from Table~\ref{tab:per_parms}. With no spiral structure/arm position \textit{a priori}, maser positions converge to a much cleaner spiral shape. We have excluded G021.87 and G070.33 from the parallax distance plot as they do not have a well-measured parallax. The thin concentric rings centered on the Sun (red star at [0,8.15]) are spaced by 1~kpc, and the thick concentric circles are spaced by 5~kpc. The rays show Galactic longitude, spaced by 10~deg. The green square shows the location of the Galactic center at $[0,0]$ in both plots.}
        \label{fig:par2ml}
    \end{figure}

    In Figure~\ref{fig:par2ml} we compare the distribution of Perseus arm distances estimated purely from parallax measurements versus combining parallax and 3DKD information as described above.  The latter method appears to naturally `tighten up' the distribution of distance points, resulting in a more visually obvious arm, and ultimately an arm fit with lower uncertainty. It is worth reiterating that no prior was used that characterizes the position or shape of the Perseus arm, and that the convergence of distances into a clear spiral arm shape arises by purely considering the astrometric measurements and the model for the Galactic rotation curve. 

\begin{table*}[ht]
    \centering
    \begin{tabular}{lcrrrr}
\toprule        \multicolumn{1}{c}{Model}  & \multicolumn{1}{c}{$\beta_k$} & \multicolumn{1}{c}{$R_k$} & \multicolumn{1}{c}{$\psi_<$} & \multicolumn{1}{c}{$\psi_>$}  & \multicolumn{1}{c}{width} \\   
\multicolumn{1}{c}{}  & \multicolumn{1}{c}{(deg)} & \multicolumn{1}{c}{(kpc)} & \multicolumn{1}{c}{(deg)} & \multicolumn{1}{c}{(deg)}  & \multicolumn{1}{c}{(kpc)} \\\midrule
        Two segment \citep{Reid2019ApJ...885..131R}  & $40$ & $8.87\pm0.13$ & $10.3\pm1.4$ &$8.7\pm2.7$ & $0.35\pm0.06$ \\
        Two segment (this work) & $40$ & $9.29\pm0.10$ & $5.9\pm1.2$ &$10.6\pm1.0$ & $0.32\pm0.04$ \\
        One segment (this work) & $40$ & $9.08\pm0.08$ & \multicolumn{2}{c}{$8.7\pm0.7$} & $0.36\pm0.06$ \\
        \bottomrule
    \end{tabular}
    \caption{Perseus arm fitting results. Columns are \textbf{(1)} model, \textbf{(2)} kink Galactocentric azimuth in deg, \textbf{(3)} kink Galactic radius in kpc, \textbf{(4)} pre-kink and \textbf{(5)} post-kink pitch angle in deg, \textbf{(6)} arm width in kpc. For the single segment fit, we have populated both pitch angle columns with the single fitted pitch angle.}
    \label{tab:per_parms}
\end{table*}

\begin{figure*}[ht]
        \centering
        \includegraphics[width=0.99\linewidth]{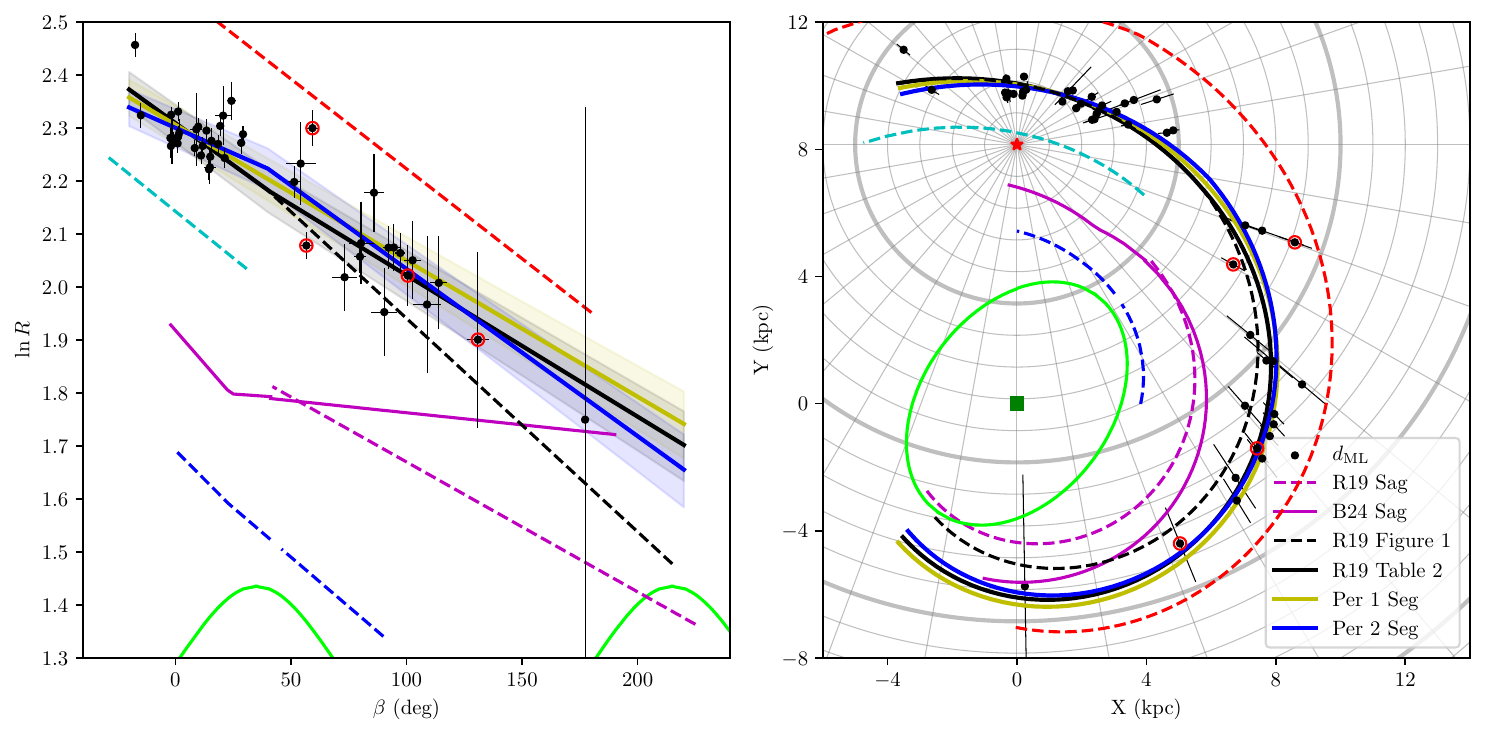}
        \caption{\textbf{Left:} The log-periodic spiral arm fits. The shaded area shows the estimated Perseus arm widths. \textbf{Right:} Plan view of the Milky Way. In both plots, the dashed lines are the spiral arm fits for the Outer (red), Perseus (black), Local (cyan), Sagittarian-Carina (magenta), and Scutum-Centaurus (blue) arms, from Table 2 \citet{Reid2019ApJ...885..131R}. The solid black line shows the Perseus arm fit from Figure 1 \citet{Reid2019ApJ...885..131R}, while the other solid lines show the new models from this work (yellow for single-segment, blue for two-segment Perseus arm), the Sagittarius arm \citep[magenta; ][]{Bian2024AJ....167..267B}, and X1 bar orbitals \citep[lime; ][]{Kumar2025ApJ...982..185K}. The green square shows the location of the Galactic center $[0,0]$, the Sun is shown with a red star at $[0,8.15]$, and new maser parallaxes from this work are highlighted with red circles.}
        \label{fig:planeview}
    \end{figure*}

    \subsection{Spiral arm fitting} \label{sec:spiral} 
    Using the measured $(l,b,d_\mathrm{ML})$ values for each maser, we project the position of each source onto the Galactic plane (see Figure~\ref{fig:par2ml}, right panel).  In order to fit spiral arm segments to these data, we use a polar Galactic coordinate system $(R,\beta)$, where $R$ is the Galactic radius and $\beta$ is the Galactic azimuth increasing clockwise from the Sun. This requires a distance to the Galactic center $R_0$, which we adopt to be $R_0=8.15$~kpc \citep{Do2019Sci...365..664D, GravCollab2019A&A...625L..10G}. The methods used to convert $(l,b,d_\mathrm{ML})\rightarrow(R,\beta)$, is given in the Appendix. We model the spiral arm as log-periodic with an optional ``kink'' \citep{Reid2019ApJ...885..131R}:
    \begin{equation}
        \ln{R} =\ln{R_k}-(\beta - \beta_k)\tan\psi
    \end{equation}  The pitch angle $\psi$ is allowed to have different values pre- and post-kink position $(\beta_k,\ln{R_k})$; $\psi = \psi_<$ if $\beta<\beta_k$, $\psi = \psi_>$ if $\beta>\beta_k$. We consider two models, one allowing for a different pitch angle about the kink and one with a single pitch angle.
    
    Using a Bayesian Markov Chain Monte Carlo approach described in \citet{Reid2014ApJ...783..130R}, we equally fit all data and the fits are given in Table~\ref{tab:per_parms}.

\section{The Perseus Arm}
    \subsection{Location and Pitch Angle} 
    In our two-segment model, the value for the post-kink pitch angle of $\psi_>=10.6\pm1.0^\circ$ is slightly larger and more precise than the previous estimate of $\psi_>=8.7\pm2.7^\circ$, but consistent within $1\sigma$. However, the greatest agreement can be found by comparing the original two-segment model from \citet{Reid2019ApJ...885..131R}, and our single-segment arm model. In general, we favor the single segment model, as it relies on fewer assumptions and parameters, while being consistent with both two-segment models. 


    \subsection{Perseus and Sagittarius arm bifurcation}  
    \citet{Xu2023ApJ...947...54X} propose that the Perseus and Sagittarius spiral arms originate as a single arm on the far side of the Galactic center and then bifurcate into two arms. This is supported by an updated model for the Sagittarius arm by \citet{Bian2024AJ....167..267B} in which that arm, starting at $\beta\approx70^\circ$, gradually moves radially outward by about 1~kpc at $\beta\approx130^\circ$, where it would intersect the Perseus arm.
    

    Note that the location of the bifurcation in \citet{Bian2024AJ....167..267B} was based on the model of the Perseus arm in Figure 1 of \citet{Reid2019ApJ...885..131R}. That model used a pitch angle of 13~deg for the post-kink Perseus arm. Had that model used the 8.7~deg pitch angle from the fit in Table 1 of \citet{Reid2019ApJ...885..131R}, the bifurcation azimuth would have been larger and consistent with our updated trace of the Perseus arm.

    We can refine the location of a bifurcation by using our more accurate Perseus arm model. By taking trial models for the Perseus and Sagittarius arms sampled according to their uncertainties and solving for the crossing points, we estimate that the two arms cross at $\beta={224^\circ}^{+48}_{-32}$, $R=5.5^{+0.3}_{-0.7}$~kpc for the single segment, or $\beta={185^\circ}^{+46}_{-31}$, $R=5.6^{+0.4}_{-0.3}$~kpc for the two segment model. As the arms approach one another, the relatively small uncertainties in the pitch angles translate to drastically different crossing points. However, our results indicate that the most likely point that the two arms cross/merge is in the far 4$^\mathrm{th}$ quadrant near $\beta=200^\circ$, rather than the far 1$^\mathrm{st}$ quadrant. It is unclear what form the arms take before this point. If the two arms have a common origin on the far side of the Galactic Bar, this parent `Perseus-Sagittarius-Carina' (PSC) arm may extend from the end of the bar at approximately $\beta=215^\circ$ \citep[i.e., ][]{Kumar2025ApJ...982..185K}. Targeting maser sources in regions between $l=340-360^\circ$ would be extremely valuable in determining where/if this PSC arm interacts with the bar. 

    
\section{Summary}
    We have conducted astrometric observations towards four masers in high mass star-forming regions in the 1st quadrant of the Milky Way (G021.87$+$00.01, G037.82$+$00.41, G060.57--00.18, and G070.29$+$01.60) and found that all four were in the Perseus spiral arm. 

    We propose and demonstrate a combined 3DKD and parallax maximum-likelihood distance estimation method to determine the distance to a maser given astrometric and kinematic parameters, which balances the low fractional uncertainty of nearby parallaxes and the relatively distance-invariant uncertainty of 3DKD. 

    With these new measurements and our revised distance–determination method, we recalculate distances to previously observed masers and re-evaluate the location and structure of the distant Perseus arm. The updated spatial distribution of masers traces a more coherent spiral pattern, and extends the range that the Perseus arm is constrained to $\beta\approx180^\circ$. In this extended geometry, the Perseus arm intersects the Sagittarius arm at $\beta \approx {200^\circ}$, favoring the bifurcation model proposed by \citet{Xu2023ApJ...947...54X}, albeit occurring in a different location. Using the combined dataset, we have now mapped the Perseus arm over its entire length where masers are currently known, and further extensions will require distances to masers $l<0^\circ$.

\appendix
    \section{Calculating Probability Distribution Functions}
    Here, we briefly discuss how we construct the distance PDFs used in these analyses. 
    We assume that the measured parallax, LSR velocity, Galactic longitude, and latitude proper motions have Gaussian probability distribution function priors centered on the measured values. The relationship between each of these and distance is:
    \begin{align*}
        \pi =&~ \frac{1}{d} \\
        V_\mathrm{lsr} =&~  \left(\Theta(R)\frac{R_0}{R} - \Theta_0\right)\sin{l} \\
        \mu_{l*} =&~ \frac{1}{4.74~d}\left(\Theta(R)\cos(l+\beta)-(\Theta_0+V_0)\cos{l}+U_0\sin{l}\right) \\ 
        \mu_{b} =&~  \frac{W_0}{4.74~d}
    \end{align*} where $\Theta(R)$ is a rotation curve at Galactic radius $R$, $\Theta_0$ is the local velocity, $V_0,U_0,W_0$ are the peculiar motions of the Sun with respect to $\Theta_0$ in the Galactic $X,Y,Z$ directions, and $\beta$ is the Galactocentric azimuth (increasing clockwise from the Sun). Galactocentric radius and azimuth depend on distance as:
    \begin{align*}
        R & = \sqrt{R_0^2 + d^2\cos^2{b} -2 R_0 d\cos{b}\cos{l}} \\
        \beta & = \arctan2\left(\frac{d\cos b\sin l}{R},\frac{R_0 - d\cos b \cos l}{R}\right)        
    \end{align*}
    For these analyses, we use the Universal Rotation Curve \citep{Persic1996MNRAS.281...27P} with $a_2,a_3,R_0,\Theta_0,V_0,U_0,W_0$ parameters from the A5 fit of \citet{Reid2019ApJ...885..131R}. 

    
    \section{Additional notes on individual masers}
    \textit{G229.57+00.15}: This water maser has a low fractional uncertainty parallax-distance \citep{Choi2014ApJ...790...99C} that the 3DKD agrees with. However, its position at 2.6 arm-widths from the center of the Perseus arm puts it midway between the Perseus and Outer arms. There is a bridge of HI at $l=230^\circ$ \citep[in the LAB data; ][]{Kalberla2005A&A...440..775K} connecting what we believe to be the Perseus and Outer arms \citep{Reid2016ApJ...823...77R}, and G229.57 may be associated with that bridge. Trial model fitting excluding G229 shows that it has little effect on the final fitted parameters. 
    
    \textit{G174.20-00.07 and G183.72-03.66}: As these masers are very close to the Galactic anti-center, and the parallax distances gives results which are very close to us \citep{Burns2017MNRAS.467.2367B, Choi2014ApJ...790...99C}, the recession velocity and proper motions are consistent with zero, and there is very little kinematic information to draw from. As such, a reasonable 3D kinematic distance for this source could not be determined. The parallax has a low fractional uncertainty and dominates the distance determination. 

    
    
    \textit{G049.41+00.32}: This water maser has a discrepancy between the 3D kinematic distance ($11.7\pm0.8$~kpc) and the parallax-distance \citep[$7.6\pm1.8$~kpc; ][]{Zhang2019AJ....157..200Z}. The original parallax has a fractional uncertainty of 23\%, giving it a skewed parallax PDF. The parallax distance would place it between the Sagittarius and Perseus arms, the velocity ($V_\mathrm{lsr}=-12\pm5$~\kms) matches the Perseus arm, so we find that the maximum likelihood distance is $11.6\pm0.6$~kpc.



    \textit{G001.00-00.23}: The original parallax measurement for this source of $0.09\pm0.06$~mas \citep{Reid2019ApJ...885..131R} has a high fractional uncertainty, serving as a lower distance limit of $\approx5$~kpc. The source is almost directly towards the Galactic Center and has almost no line-of-sight velocity component. Therefore, the maximum likelihood distance is based almost solely on the parallax and the longitude proper motion, which places it at $d_\mathrm{ML}=13.9^{+3.3}_{-3.5}$~kpc, and very close to the bifurcation region. 


    \begin{table*}[h]
\footnotesize
        \centering
		\begin{tabular}{lrrrrrrcl}
			\toprule
\multicolumn{1}{c}{\bf Maser Name}&\multicolumn{1}{c}{$\boldsymbol{\pi}$}&\multicolumn{1}{c}{$\boldsymbol{\mu_x}$}&\multicolumn{1}{c}{$\boldsymbol{\mu_y}$}&\multicolumn{1}{c}{$\boldsymbol{V_\mathrm{lsr}}$}&\multicolumn{1}{c}{$\boldsymbol{D_\pi}$}&\multicolumn{1}{c}{$\boldsymbol{d_\mathrm{3Dk}}$}&\multicolumn{1}{c}{$\boldsymbol{d_\mathrm{ML}}$}&\multicolumn{1}{l}{Ref$^\dagger$}\\
\multicolumn{1}{c}{}&\multicolumn{1}{c}{(mas)}&\multicolumn{1}{c}{(mas/yr)}&\multicolumn{1}{c}{(mas/yr)}&\multicolumn{1}{c}{(\kms)}&\multicolumn{1}{c}{(kpc)}&\multicolumn{1}{c}{(kpc)}&\multicolumn{1}{c}{(kpc)}&\multicolumn{1}{c}{}\\ \midrule
G001.00$-$00.23(C)   & $0.090\pm0.057$ & $-3.87\pm 0.28$ & $-6.23\pm0.56$ & $ +2.0\pm 5.0 $ & $11.10_{-3.47}^{+8.70}$ & $13.95_{-5.76}^{+3.62}$ & $13.90_{-3.50}^{+3.29}$ & 24 \\ 
G021.87$+$00.01(H)   & $0.174\pm0.126$ & $-3.17\pm 0.48$ & $-6.13\pm0.21$ & $+19.0\pm10.0 $ & $ 5.75_{-2.33}^{+11.38}$ & $13.55_{-1.44}^{+1.51}$ & $13.53_{-1.20}^{+1.28}$  & 1 \\ 
G031.24$-$00.11(H)   & $0.076\pm0.014$ & $-2.80\pm 0.19$ & $-5.54\pm0.19$ & $+24.0\pm10.0 $ & $13.15_{-2.21}^{+3.32}$ & $13.10_{-1.78}^{+2.19}$ & $13.10_{-0.78}^{+0.80}$ & 2\\ 
G032.79$+$00.19(H)   & $0.103\pm0.031$ & $-2.94\pm 0.24$ & $-6.07\pm0.25$ & $+16.0\pm10.0 $ & $ 9.70_{-2.60}^{+5.38}$ & $12.53_{-2.81}^{+1.71}$ & $12.47_{-1.24}^{+1.12}$ & 2\\ 
G037.49$+$00.52(H)   & $0.091\pm0.016$ & $-2.74\pm 0.27$ & $-5.49\pm0.22$ & $+11.0\pm10.0 $ & $11.00_{-1.83}^{+2.62}$ & $12.57_{-1.31}^{+1.40}$ & $12.45_{-0.74}^{+0.74}$ & 8 \\ 
G037.82$+$00.41(H)   & $0.085\pm0.007$ & $-2.68\pm 0.21$ & $-5.60\pm0.23$ & $+16.0\pm10.0 $ & $11.78_{-0.95}^{+1.04}$ & $12.22_{-1.36}^{+1.46}$ & $12.10_{-0.59}^{+0.56}$ & 1, 8 \\ 
G040.42$+$00.70(C)   & $0.078\pm0.013$ & $-2.95\pm 0.09$ & $-5.48\pm0.10$ & $+10.0\pm 5.0 $ & $12.82_{-2.00}^{+2.80}$ & $12.00_{-0.59}^{+0.60}$ & $12.05_{-0.43}^{+0.38}$ & 2\\ 
G040.62$-$00.13(C)   & $0.080\pm0.021$ & $-2.69\pm 0.12$ & $-5.60\pm0.25$ & $+31.0\pm 5.0 $ & $12.50_{-2.80}^{+4.93}$ & $10.78_{-1.92}^{+3.46}$ & $10.82_{-0.79}^{+1.01}$ & 2\\ 
G042.03$+$00.19(C)   & $0.071\pm0.012$ & $-2.40\pm 0.09$ & $-5.64\pm0.11$ & $+12.0\pm 5.0 $ & $14.07_{-2.19}^{+3.11}$ & $11.75_{-1.05}^{+1.08}$ & $11.85_{-0.48}^{+0.49}$ & 2\\ 
G043.16$+$00.01(H)   & $0.090\pm0.005$ & $-2.48\pm 0.15$ & $-5.27\pm0.13$ & $+11.0\pm 5.0 $ & $11.38_{-0.65}^{+0.66}$ & $11.78_{-3.23}^{+4.48}$ & $11.62_{-0.48}^{+0.42}$ & 3\\ 
G048.60$+$00.02(H)   & $0.093\pm0.005$ & $-2.89\pm 0.13$ & $-5.50\pm0.13$ & $+18.0\pm 5.0 $ & $10.75_{-0.55}^{+0.62}$ & $ 9.93_{-1.68}^{+1.57}$ & $10.28_{-0.37}^{+0.39}$ & 3\\ 
G049.26$+$00.31(C)   & $0.113\pm0.016$ & $-2.73\pm 0.15$ & $-5.85\pm0.19$ & $ +0.0\pm 5.0 $ & $ 8.85_{-1.18}^{+1.56}$ & $10.60_{-2.92}^{+1.06}$ & $10.45_{-1.16}^{+0.78}$ & 2\\ 
G049.41$+$00.32(C)   & $0.132\pm0.031$ & $-3.15\pm 0.17$ & $-4.49\pm0.66$ & $-12.0\pm 5.0 $ & $ 7.58_{-1.71}^{+3.01}$ & $11.70_{-1.39}^{+1.80}$ & $11.60_{-0.92}^{+0.91}$ & 2\\ 
G050.28$-$00.39(H)   & $0.135\pm0.026$ & $-3.29\pm 0.56$ & $-5.52\pm0.37$ & $+17.0\pm 5.0 $ & $ 7.40_{-1.34}^{+2.08}$ & $ 9.50_{-4.02}^{+1.59}$ & $ 9.38_{-0.93}^{+0.78}$ & 9 \\  
G060.57$-$00.18(C)   & $0.130\pm0.009$ & $-3.23\pm 0.13$ & $-5.68\pm0.13$ & $ +3.0\pm 5.0 $ & $ 7.70_{-0.53}^{+0.55}$ & $ 7.67_{-4.10}^{+1.71}$ & $ 7.67_{-0.41}^{+0.40}$ & 1, 24 \\ 
G070.18$+$01.74(C)   & $0.136\pm0.014$ & $-2.88\pm 0.15$ & $-5.18\pm0.18$ & $-23.0\pm 5.0 $ & $ 7.35_{-0.71}^{+0.88}$ & $ 7.60_{-1.43}^{+1.21}$ & $ 7.50_{-0.48}^{+0.50}$ & 2\\ 
G070.29$+$01.60(H)   & $0.103\pm0.008$ & $-2.81\pm 0.20$ & $-4.44\pm0.20$ & $-27.0\pm10.0 $ & $ 9.70_{-0.73}^{+0.82}$ & $ 8.35_{-1.17}^{+1.12}$ & $ 9.12_{-0.51}^{+0.54}$ & 1 \\ 
G070.33$+$01.58(H)   & $0.074\pm0.037$ & $-2.82\pm 0.29$ & $-4.68\pm0.28$ & $-23.0\pm 5.0 $ & $13.53_{-3.59}^{+7.23}$ & $ 7.88_{-0.99}^{+0.93}$ & $ 8.05_{-1.17}^{+1.36}$ & 9 \\  
G094.60$-$01.79(C,H) & $0.221\pm0.013$ & $-3.49\pm 0.23$ & $-2.65\pm0.23$ & $-43.0\pm 5.0 $ & $ 4.53_{-0.26}^{+0.28}$ & $ 5.25_{-0.82}^{+0.79}$ & $ 4.65_{-0.27}^{+0.27}$ & 4, 5 \\ 
G095.29$-$00.93(C,H) & $0.206\pm0.007$ & $-2.74\pm 0.30$ & $-2.85\pm0.37$ & $-38.0\pm 5.0 $ & $ 4.85_{-0.17}^{+0.17}$ & $ 4.92_{-1.14}^{+1.17}$ & $ 4.85_{-0.15}^{+0.17}$ & 4, 15  \\ 
G100.37$-$03.57(H)   & $0.289\pm0.016$ & $-3.66\pm 0.60$ & $-3.02\pm0.60$ & $-37.0\pm10.0 $ & $ 3.45_{-0.18}^{+0.21}$ & $ 4.15_{-1.30}^{+1.29}$ & $ 3.50_{-0.20}^{+0.19}$ & 4 \\ 
G108.20$+$00.58(H)   & $0.227\pm0.037$ & $-2.25\pm 0.50$ & $-1.00\pm0.50$ & $-49.0\pm 5.0 $ & $ 4.40_{-0.68}^{+0.95}$ & $ 4.67_{-1.52}^{+6.77}$ & $ 4.55_{-0.49}^{+0.54}$ & 4 \\ 
G108.42$+$00.89(H)   & $0.400\pm0.050$ & $-2.58\pm 0.60$ & $-1.91\pm0.50$ & $-51.0\pm 5.0 $ & $ 2.50_{-0.30}^{+0.36}$ & $ 4.75_{-1.08}^{+1.10}$ & $ 2.52_{-0.35}^{+0.58}$ & 16 \\ 
G108.47$-$02.81(H)   & $0.309\pm0.010$ & $-3.13\pm 0.49$ & $-2.79\pm0.46$ & $-54.0\pm 5.0 $ & $ 3.23_{-0.10}^{+0.11}$ & $ 4.80_{-1.05}^{+1.04}$ & $ 3.25_{-0.12}^{+0.10}$ & 4 \\ 
G108.59$+$00.49(H)   & $0.405\pm0.033$ & $-5.56\pm 0.40$ & $-3.40\pm0.40$ & $-52.0\pm 5.0 $ & $ 2.48_{-0.20}^{+0.21}$ & $ 4.83_{-4.79}^{+1.08}$ & $ 2.45_{-0.20}^{+0.23}$ & 4 \\ 
G111.23$-$01.23(H)   & $0.300\pm0.081$ & $-4.37\pm 0.60$ & $-2.38\pm0.60$ & $-53.0\pm10.0 $ & $ 3.33_{-0.92}^{+1.98}$ & $ 4.28_{-1.44}^{+1.45}$ & $ 3.88_{-0.78}^{+0.87}$ & 4 \\ 
G111.25$-$00.76(C,H) & $0.280\pm0.015$ & $-2.69\pm 0.28$ & $-1.75\pm0.33$ & $-40.0\pm 3.0 $ & $ 3.58_{-0.19}^{+0.20}$ & $ 3.67_{-0.65}^{+0.67}$ & $ 3.58_{-0.17}^{+0.20}$ & 4, 20 \\ 
G111.54$+$00.77(C)   & $0.378\pm0.017$ & $-2.45\pm 0.40$ & $-2.44\pm0.40$ & $-57.0\pm 5.0 $ & $ 2.65_{-0.12}^{+0.12}$ & $ 4.95_{-1.01}^{+1.05}$ & $ 2.65_{-0.13}^{+0.11}$ & 12  \\ 
G115.05$-$00.04(H)   & $0.356\pm0.040$ & $-3.70\pm 0.78$ & $-2.00\pm0.81$ & $-36.0\pm 5.0 $ & $ 2.80_{-0.29}^{+0.37}$ & $ 3.15_{-0.83}^{+0.84}$ & $ 2.90_{-0.30}^{+0.30}$ & 11 \\ 
G122.01$-$07.08(H)   & $0.460\pm0.020$ & $-3.70\pm 0.50$ & $-1.25\pm0.50$ & $-50.0\pm 5.0 $ & $ 2.17_{-0.10}^{+0.09}$ & $ 3.88_{-1.16}^{+1.11}$ & $ 2.17_{-0.10}^{+0.09}$ & 6 \\ 
G123.06$-$06.30(C)   & $0.421\pm0.022$ & $-2.69\pm 0.31$ & $-1.77\pm0.29$ & $-29.0\pm 3.0 $ & $ 2.38_{-0.13}^{+0.12}$ & $ 2.38_{-0.52}^{+0.52}$ & $ 2.38_{-0.12}^{+0.12}$ & 22 \\ 
G123.06$-$06.31(H)   & $0.355\pm0.030$ & $-2.79\pm 0.62$ & $-2.14\pm0.70$ & $-30.0\pm 5.0 $ & $ 2.83_{-0.25}^{+0.24}$ & $ 2.50_{-0.73}^{+0.79}$ & $ 2.77_{-0.21}^{+0.22}$ & 19 \\ 
G133.94$+$01.06(C,H) & $0.511\pm0.008$ & $-1.20\pm 0.32$ & $-0.15\pm0.32$ & $-45.0\pm 3.0 $ & $ 1.95_{-0.05}^{+0.02}$ & $ 3.90_{-0.94}^{+1.06}$ & $ 1.95_{-0.05}^{+0.02}$ & 14,10,8 \\ 
G134.62$-$02.19(H)   & $0.413\pm0.017$ & $-0.49\pm 0.50$ & $-1.19\pm0.48$ & $-39.0\pm 5.0 $ & $ 2.42_{-0.10}^{+0.10}$ & $ 3.15_{-2.08}^{+10.93}$ & $ 2.42_{-0.10}^{+0.11}$  & 18 \\ 
G136.84$+$01.16(C)   & $0.442\pm0.123$ & $+0.42\pm 0.50$ & $-0.45\pm0.50$ & $-42.0\pm 5.0 $ & $ 2.27_{-0.68}^{+1.54}$ & $ 1.38_{-0.69}^{+3.43}$ & $ 2.30_{-0.57}^{+1.04}$ & 20     \\ 
G170.65$-$00.24(H)   & $0.532\pm0.075$ & $+0.23\pm 1.20$ & $-3.14\pm0.60$ & $-19.0\pm 5.0 $ & $ 1.88_{-0.25}^{+0.33}$ & $ 1.25_{-0.95}^{+10.66}$ & $ 1.75_{-0.18}^{+0.25}$  & 7 \\ 
G173.17$+$02.36(H)   & $0.590\pm0.043$ & $+0.01\pm 1.20$ & $-3.40\pm1.20$ & $-10.0\pm10.0 $ & $ 1.70_{-0.13}^{+0.12}$ & $ 1.15_{-0.95}^{+7.30}$ & $ 1.68_{-0.13}^{+0.11}$ & 28 \\ 
G173.48$+$02.44(C)   & $0.594\pm0.014$ & $+0.62\pm 0.63$ & $-2.34\pm0.63$ & $-13.0\pm 5.0 $ & $ 1.68_{-0.04}^{+0.05}$ & $ 1.52_{-1.17}^{+10.19}$ & $ 1.68_{-0.03}^{+0.05}$  & 20 \\ 
G173.71$+$02.69(H)   & $0.639\pm0.033$ & $+0.79\pm 1.25$ & $-2.41\pm1.25$ & $-17.0\pm10.0 $ & $ 1.57_{-0.09}^{+0.07}$ & $ 1.40_{-1.12}^{+10.87}$ & $ 1.55_{-0.07}^{+0.09}$  & 26 \\  
G174.20$-$00.07(H)   & $0.467\pm0.020$ & $+0.32\pm 0.56$ & $-0.22\pm0.68$ & $ -2.0\pm10.0 $ & $ 2.15_{-0.11}^{+0.08}$ & $11.22_{-5.05}^{+10.19}$ & $ 2.15_{-0.10}^{+0.08}$  & 25 \\ 
G183.72$-$03.66(H)   & $0.629\pm0.012$ & $+0.38\pm 1.30$ & $-1.32\pm1.30$ & $ +3.0\pm 5.0 $ & $ 1.60_{-0.05}^{+0.02}$ & $ 2.65_{-2.03}^{+8.68}$ & $ 1.60_{-0.05}^{+0.02}$ & 4 \\ 
G188.79$+$01.03(H)   & $0.496\pm0.103$ & $-0.10\pm 0.50$ & $-3.91\pm0.50$ & $ +1.0\pm 5.0 $ & $ 2.02_{-0.41}^{+0.64}$ & $ 0.82_{-0.41}^{+0.39}$ & $ 1.62_{-0.25}^{+0.60}$ & 21 \\ 
G188.94$+$00.88(C,H) & $0.476\pm0.006$ & $-0.32\pm 0.49$ & $-1.94\pm0.49$ & $ +8.0\pm 5.0 $ & $ 2.10_{-0.03}^{+0.03}$ & $ 1.93_{-1.08}^{+1.38}$ & $ 2.10_{-0.03}^{+0.03}$ & 5, 13, 20 \\ 
G192.16$-$03.81(C,H) & $0.659\pm0.070$ & $+0.69\pm 0.70$ & $-1.57\pm0.70$ & $ +5.0\pm 5.0 $ & $ 1.52_{-0.16}^{+0.18}$ & $ 1.45_{-0.77}^{+0.94}$ & $ 1.52_{-0.16}^{+0.16}$ & 22, 27  \\ 
G192.60$-$00.04(H)   & $0.601\pm0.039$ & $-0.18\pm 0.60$ & $-0.43\pm0.60$ & $ +7.0\pm 5.0 $ & $ 1.68_{-0.12}^{+0.10}$ & $ 2.08_{-2.05}^{+11.45}$ & $ 1.68_{-0.12}^{+0.11}$  & 17 \\ 
G229.57$+$00.15(H)   & $0.218\pm0.012$ & $-1.33\pm 0.70$ & $+0.77\pm0.70$ & $+53.0\pm10.0 $ & $ 4.58_{-0.23}^{+0.28}$ & $ 5.05_{-2.30}^{+5.53}$ & $ 4.60_{-0.24}^{+0.26}$ & 4 \\  
G236.81$+$01.98(H)   & $0.326\pm0.026$ & $-2.49\pm 0.43$ & $+2.67\pm0.41$ & $+53.0\pm 5.0 $ & $ 3.08_{-0.25}^{+0.25}$ & $ 3.58_{-1.63}^{+1.47}$ & $ 3.15_{-0.26}^{+0.25}$ & 4 \\ \hline
        \end{tabular}
        \caption{Astrometric and kinematic properties of other candidate Perseus arm masers. Columns are: (1) Maser name (type of maser C=methanol, H=water), (2) parallax, (3) proper motion in $\alpha\cos\delta$, (4) and $\delta$ directions, (5) LSR velocity, (6) parallax-distance, (7) 3D kinematic distance, (8) maximum likelihood distance and, (9) references. When there are multiple parallaxes and proper motions measured, we report the variance-weighted parallax in columns 2-4. For the three distance estimates, we report the mode of the probability distribution, and the smallest upper and lower limits that enclose 68.3\% confidence (to account for skewed distributions). $\dagger$Ref- 1: This work, 2: \citet{Zhang2019AJ....157..200Z}, 3: \citet{Zhang2013ApJ...775...79Z}, 4: \citet{Choi2014ApJ...790...99C}, 5: \citet{Oh2010PASJ...62..101O}, 6: \citet{Moellenbrock2009ApJ...694..192M}, 7: \citet{Sakai2012PASJ...64..108S}, 8: \citet{Nagayama2020PASJ...72...52N}, 9: \citet{Sakai2022PASJ...74..209S}, 10: \citet{Matsumoto2011PASJ...63.1345M}, 11: \citet{Kusuno2013ApJ...774..107K}, 12: \citet{Moscadelli2009ApJ...693..406M}, 13: \citet{Reid2009ApJ...693..397R}, 14: \citet{Xu2006Sci...311...54X}, 15: \citet{Nakanishi2015PASJ...67...68N}, 16: \citet{Imai2012PASJ...64..142I}, 17: \citet{Shiozaki2011PASJ...63.1219S}, 18: \citet{Asaki2010ApJ...721..267A}, 19: \citet{Sato2008PASJ...60..975S}, 20: \citet{Sakai2019ApJ...876...30S}, 21: \citet{Niinuma2011PASJ...63....9N}, 22: \citet{Rygl2010A&A...511A...2R}, 24: \citet{Reid2019ApJ...885..131R}, 25: \citet{Burns2017MNRAS.467.2367B}, 26: \citet{Burns2015MNRAS.453.3163B}, 27: \citet{Burns2016MNRAS.460..283B}, 28: \citet{Vera2020PASJ...72...50V}. 
        }
    \label{tab:allmasers}
    \end{table*}

\bibliography{perseus}
\bibliographystyle{aasjournal}
\end{document}